\documentstyle[preprint,aps]{revtex}
\tighten

\begin{document}
\draft

\title{Examination of Wandzura-Wilczek Relation \\
	for $g_2(x,Q^2)$ in pQCD}
\author{{\bf A. Harindranath$^a$ and Wei-Min Zhang$^b$} \\
	$^a$Saha Institute of Nuclear Physics, 1/AF Bidhan Nagar, 
		Calcutta, 700064 India \\
	$^b$Institute of Physics, Academia Sinica, Taipei, 11529 Taiwan, 
		R.O.C.}
\date{March 14, 1997, Revised May 20, 1997}
\maketitle

\begin{abstract}
In order to examine the validity of Wandzura-Wilczek relation 
for the polarized DIS structure function $g_2(x,Q^2)$, we use 
the light-front time-ordering pQCD to calculate $g_2(x,Q^2)$ at 
order $\alpha_s$ on a quark target. We find that the study of 
the transverse polarized structure function in pQCD is only 
meaningful if we begin with massive quarks. The result shows 
that the Wandzura-Wilczek relation for $g_2(x,Q^2)$ is strongly 
violated by the quark mass in pQCD.
\end{abstract}
\centerline{PACS: 11.10Ef; 11.40.-q; 12.38.Bx}

\vspace{0.5in}

\noindent{\bf 1. Introduction}

The transverse polarized structure function $g_2(x,Q^2)$ is perhaps 
the least well-known structure function in deep inelastic 
lepton-nucleon scatterings (DIS). In the early naive parton model,
Feynman claimed that $g_2$, just like $g_1$, has a simple
parton interpretation: $g_1(x) + g_2(x) = {1\over 2} \sum_q e_q^2 
\Delta q_T(x)$, where $\Delta q_T(x)$ is the difference between the
number density of quarks polarized along the same direction of the
transverse polarization of nucleon and these polarized along the 
opposite direction \cite{Feynman72}. He also pointed out that
one would permit a measurement of the transverse quark polarization 
if one keeps terms proportional to quark masses. This implies that 
the measurement of transverse polarization in DIS is a ``higher 
twist'' effect, based on some common understanding of the concept 
of twist. Some years later, Wandzura and Wilczek studied the properties
of $g_2$ in DIS by the use of the operator product expansion (OPE)
analysis \cite{Wandzura77}. They claimed that except for a twist
three contribution $\overline{g}_2(x)$ which may be negligible in 
their model calculation, $g_2(x)$ can be related to an integral 
over $g_1(x)$:
\begin{equation}
	g_2(x) = g_2^{WW}(x) + \overline{g}_2(x), ~~~~~~~~
	g_2^{WW}(x) = - g_1(x) + \int_x^1 {dy\over y} g_1(y) .
\end{equation}
The relation between $g_2^{WW}$ with $g_1(x)$ is called 
Wandzura-Wilczek (WW) relation. In their work, the quark mass effect
is neglected.  After a couple of years, Altarelli and Muzzetto explicitly 
computed $g_2(x,Q^2)$ in perturbative QCD and found that quark masses 
play an important and nontrivial role in determining $g_2$ but 
unfortunately the work was not published \cite{Altarelli79}. 
The OPE analysis with the quark mass effect included was given 
by Kodaira et al. \cite{Kodaira79} at the same time  
which results in the same solution as in \cite{Altarelli79} but the 
picture of $g_2$ is not very clear in OPE analysis. 
Later 
on, Shuryak and Vainshtein pointed out that the twist-three 
contribution $\overline{g}_2$ is a direct quark-gluon interaction 
effect which is important in $g_2(x,Q^2)$ \cite{Chuyk84} . Thus, 
the measurement of $g_2$ may also be very sensitive to the 
interaction dependent higher twist effects in QCD \cite{Jaffe90}. 

Since then, much work on the subject is concentrated on the 
questions whether $g_2$ can be described approximately by 
Feynman's parton picture \cite{Ralston,Leader,Man}, and whether 
it is a relatively good approximation to predict $g_2$ from the
longitudinal polarized structure function $g_1$ via the WW relation 
\cite{Roberts89,Roberts95}, or whether the quark-gluon coupling 
(a twist-3 operator) can provide a significant contribution to $g_2$ 
\cite{Jaffe91,Strat,Song}. Different authors used different methods 
and obtained different results, and these are not all compatible 
with each other\cite{Jaffe90,Review}. However, because of the lack 
of experimental data, the understanding of $g_2$ is very limited 
in the investigations of past twenty years. Very recently, 
$g_2(x,Q^2)$ has been preliminarily measured by the SMC experiment 
in CERN and the E143 experiment in SLAC \cite{E143}. Although it 
seems that the existing data can be fit quite well by the WW 
relation, whether this simple relation can reasonably describe 
the transverse polarization physics of hadrons in DIS and whether 
one may also see the effect of direct quark-gluon interaction 
from $g_2$ are still in question. 

Roberts and Ross attempted to show from their parton model picture 
that the WW relation should be a good approximation in the description 
of $g_2$ \cite{Roberts95} even if quark masses and quark transverse 
momentum are important in determining $g_2$. While, Jaffe and Ji 
have argued from OPE that the twist-3 contribution $\bar{g}_2$ may be 
very significant \cite{Jaffe91}.
Some bag model calculations have demonstrated that the twist-3 
contribution is significantly large in comparison with $g_2^{WW}$ 
\cite{Strat,Song}. The preliminary data is not sufficient to justify 
these arguments.

Physically, if the WW relation would be a good description of $g_2$, 
then $g_2$ would provide no new information about QCD dynamics in 
transversely polarized hadrons since $g_2$ can be determined from $g_1$. 
This seems to be highly unlikely.  On the other hand, if the twist-3 
contribution in $g_2$ would be significantly large and, as the bag 
model calculations claimed\cite{Jaffe91,Strat,Song} there should be 
a huge cancellation between the twist-3 and the twist-2 contributions 
for $g_2$, then the classification in terms of twist may not 
be useful. This certainly contradicts the current understanding 
of DIS phenomena based on the twist expansion in $1/Q$.

Indeed, to get an intuitive picture of deep inelastic scattering in
field theory, it is extremely helpful to keep close contact with
parton ideas. However, partons were originally introduced as 
collinear, massless, on-mass shell objects which do not interact
with each other. The question, then, is if one can generalize this 
concept and introduce {\it field theoretic partons} as non-collinear
and massive (in the case of quarks) but still on-mass shell objects 
in interacting field theory? Light-front Hamiltonian description of 
composite systems which utilizes many body wavefunctions for the 
constituents allows us to precisely achieve this goal.  

Within the light-front Hamiltonian description, we recently
find that the pure transverse polarized structure function in DIS, 
\begin{equation}  \label{gtg2}
	g_T(x,Q^2)=g_1(x,Q^2)+g_2(x,Q^2),
\end{equation}
constitutes indeed a direct measurement of the QCD dynamics of chiral 
symmetry breaking\cite{Zhang96}. The major contributions from all the 
sources (the quark mass effect, the quark transverse momentum 
contribution and the direct quark-gluon coupling dynamics) to $g_T$ 
are proportional to the simple transverse polarized parton distribution 
\cite{Feynman72} due to the dynamical chiral symmetry breaking.  Based 
on the hadronic bound state structure analysis, the so-called truly 
twist-3 contributions from the direct quark-gluon coupling and transverse 
quark momentum dynamics which have no simple parton interpretation 
should be rather unimportant. Our result indicates that the WW relation 
is not essential in determining $g_2$.

In this paper we shall compute the transverse polarized structure 
function in light-front Hamiltonian (namely time-ordering) perturbative 
QCD (pQCD) up to the order of $\alpha_s$ on a quark target. The 
result explicitly shows that the WW relation is 
{\it strongly} violated in pQCD by the quark mass effect (or chiral 
symmetry breaking dynamics). 

\vspace{0.3in}

\noindent{\bf 2. Light-front time-ordered pQCD calculation of $g_2(x,Q^2)$}

Recently, several authors have used different methods to calculate 
$g_2$ on the quark target to examine if it obeys the so-called 
Burkhardt-Cottingham (BC) sum rule \cite{BC} in pQCD \cite{Ven93,Alta94}.
We find that BC sum rule is a trivial result in our formulation 
to all orders in pQCD (see later discussion). But so far, none has
explicitly paid attention on the WW relation in pQCD. Of course, 
the result on quark target cannot be tested from experiments, but 
it can provide some basis for the theoretical understanding of the 
polarized structure functions themselves. This is not only because 
the structure functions on quark target can be exactly calculated 
in pQCD, but with the use of the factorization theorem, all we have 
understood from QCD about the hard processes extracted from DIS are 
also indeed based on these quark and gluon target structure function 
calculations \cite{pQCD}. Here we shall use the light-front Hamiltonian 
analysis of DIS structure functions \cite{Hari96} to directly calculate 
$g_1(x,Q^2)$ and $g_T(x,Q^2)$ in pQCD. In light-front Hamiltonian 
approach (utilizing old-fashioned perturbation theory) the physical 
picture is very transparent at each stage of the calculation since all 
the calculations within this framework are performed in the physical space. 
In addition, compared to the other approaches\cite{Ven93,Alta94}, the 
light-front formulation calculations are far more simple and straight 
forward. 

Note that both in experiment and theory, with the target 
being polarized either in the longitudinal or transverse direction, 
one can only measure and calculate either the helicity structure 
function $g_1$ or the transverse polarized structure function $g_T$.  
The $g_2$ structure function which is well-defined from the hadronic 
tensor based on the Lorentz structure and the current conservation, 
however, cannot be measured and computed directly. It can only be 
extracted from the definition eq.(\ref{gtg2}). This is very much
different from the calculation in the OPE framework\cite{Ven93,Alta94}.
Therefore, we shall first calculate $g_1$ and $g_T$ in the quark 
helicity state and the quark state polarized in transverse direction, 
respectively, up to $\alpha_s$ in pQCD. Then we can obtain the 
$g_2$ unambiguously from eq.(\ref{gtg2}).

As we have recently shown\cite{Zhang96}, in the large $q^-$ (BJL) 
limit\cite{Jackiw}, the leading contributions to the polarized 
structure functions in DIS are given by
\begin{eqnarray}
	&& g_1(x,Q^2) = {1\over 8\pi S^+} \int_{-\infty}^{\infty}
		d\eta e^{-i\eta x} \langle PS |~ \overline{\psi}(\xi^-)
		{\cal Q}^2 \gamma^+ \gamma_5 \psi(0)+ h.c. |PS \rangle, 
		\label{g1} \\
	&& g_T(x,Q^2) = {1\over 8\pi (S_\bot-{P_\bot\over P^+}S^+) } 
		\int_{-\infty}^{\infty} d\eta e^{-i\eta x} \langle PS 
		|~ \overline{\psi}(\xi^-){\cal Q}^2 \nonumber \\
	&& ~~~~~~~~~~~~~~~~~~~~~~~~~~~~~~~~~~~~~ \times 
		\Bigg(\gamma_\bot - {P_\bot \over P^+ }\gamma^+ 
		\Bigg)\gamma_5 \psi(0) + h. c. |PS \rangle , \label{gt} 
\end{eqnarray}
where $P$ and $S$ are the target four-momentum and polarization 
vector, respectively ($P^2=M^2, S^2=-M^2, S \cdot P=0$), and $q$ is 
the virtual-photon four momentum ($Q^2=-q^2$, $\nu = P\cdot q$, 
$x=Q^2/2\nu$).    The parameter $\eta \equiv {1\over 2}P^+ \xi^-$ 
with $\xi^-$ being the light-front longitudinal coordinate, and 
${\cal Q}$ the quark charge operator.  This is a general expression 
for the DIS polarized structure functions in which the target has 
not been assigned to any specific Lorentz frame. If we let the target 
be in the rest frame, eqs.(\ref{g1}-\ref{gt}) are reduced to the 
expressions previously derived by Jaffe and Ji in the impulse 
approximation \cite{Jaffe91} and also by Efremov et al. in QCD 
field theoretic model \cite{Ef84}.

From eqs.(\ref{g1}) and (\ref{gt}), one can easily obtain
\begin{eqnarray}
	g_2(x,Q^2) &=& {1\over 8\pi(S_\bot-{P_\bot\over P^+})}
		\int_{-\infty}^{\infty} d\eta e^{-i\eta x} \langle PS 
		|~ \overline{\psi}(\xi^-){\cal Q}^2 \nonumber \\
	&& ~~~~~~~~~~~~~~~~~~~~~~~~~~~~~~~~~~~~~ \times 
		\Bigg(\gamma_\bot - {S_\bot \over S^+ }\gamma^+ 
		\Bigg)\gamma_5 \psi(0) + h. c. |PS \rangle , \label{g2} 
\end{eqnarray}		
This is again the most general matrix element expression for
$g_2(x,Q^2)$. Note that if there is no anomalous contribution,
\begin{equation}
	\langle PS |~ \overline{\psi}_i \gamma^\mu \gamma_5
		\psi_i |PS \rangle ~ \sim ~ \Delta q_i S^\mu .
		\label{gme} 
\end{equation}
Then it is straightfroward to show that
\begin{equation}
	\int_0^1 dx g_2(x,Q^2) = {1\over 2(S_\bot-{P_\bot\over P^+})}
		\langle PS |~ \overline{\psi}(0){\cal Q}^2
		\Bigg(\gamma_\bot - {S_\bot \over S^+ }\gamma^+ 
		\Bigg)\gamma_5 \psi(0)|PS \rangle = 0
\end{equation}
which is just the BC sum rule and it is valid to all orders in pQCD 
and in the full theory of hadron dynamics. But anomalous contribution
may spoil this sum rule which we will not discuss in this paper.

Since we will focus on the WW relation in $g_2(x,Q^2)$ in this
paper, we must individually calculate the contributions of
$g_2(x,Q^2)$ from various sources: 
the quark mass contribution, the quark transverse 
momentum contribution and quark-gluon coupling effect, etc.
It is perhaps the most convenient to perform such calculation
in the framework of light-front field theory. As we can see, 
the matrix elements in 
eqs.({\ref{g1}-\ref{gt}) are expressed on the equal light-front 
time surface.  On the light-front, $\psi$ is decoupled into 
$\psi = \psi_+ + \psi_-$, $\psi_\pm = {1\over 2} \gamma^0 \gamma^+ 
\psi$, and the component $\psi_-$ is not physically independent: 
$\psi_- = {\gamma^0 \over i\partial^+}(i{\not \! \! D_\bot} +m)
\psi_+$ in QCD, where $D_\bot = \partial_\bot -igA_\bot$ is
the transverse component of the covariant derivative \cite{Zhang93}. 
Hence the precise expression of  eqs.(\ref{g1}-\ref{gt}) should 
be\cite{Zhang96} \begin{eqnarray}
	g_1(x,Q^2) &=& {1\over 4\pi S^+} \int_{-\infty}^{\infty}
		d\eta e^{-i\eta x} \langle P S| \psi_+^\dagger (\xi^-)
		{\cal Q}^2 \gamma_5 \psi_+(0) + h.c. |PS \rangle , 
		\label{g1lf} \\
	g_T(x,Q^2) &=& {1\over 8\pi \left(S_\bot-{P_\bot\over
                                 P^+}S^+ \right)} 
		\int_{-\infty}^{\infty} 
		d\eta e^{-i\eta x} \langle PS | \Big(O_m + O_{k_\bot} 
		+ O_g \Big)  + h.c |PS \rangle  \nonumber \\
		&=& g_T^m(x,Q^2) + g_T^{k_\bot}(x,Q^2) + g_T^g
			(x,Q^2),  \label{gtlf} 
\end{eqnarray}
where 
\begin{eqnarray}
	&& O_m=m_q \psi_+^\dagger (\xi^-) {\cal Q}^2 \gamma_\bot 
		\Bigg({1 \over i \roarrow{\partial}^+} - {1\over 
		i \loarrow{\partial}^+} \Bigg)\gamma_5 \psi_+(0),
			\nonumber \\
	&& O_{k_\bot}= -\psi_+^\dagger (\xi^-){\cal Q}^2\Bigg(\gamma_\bot
		{1\over\roarrow{\partial}^+}{\not \! \roarrow{\partial_\bot}} 
		+ {\not \! \loarrow{\partial_\bot}}{1\over \loarrow{
		\partial}^+}\gamma_\bot + 2{P_\bot \over P^+}\Bigg) 
		\gamma_5\psi_+(0), \nonumber \\
	&& O_g = g\psi_+^\dagger(\xi^-){\cal Q}^2 \Bigg({\not \! \!
		A_\bot}(\xi^-){1\over i\loarrow{\partial}^+}\gamma_\bot 
		- \gamma_\bot {1\over i\roarrow{\partial}^+}{\not \! \! 
		A_\bot}(0) \Bigg)\gamma_5 \psi_+(0) ~, \label{gto}
\end{eqnarray} 
and $m_q$ and $g$ are the quark mass and quark-gluon coupling constant 
in QCD and $A_\bot=\sum_a T^a A_{a\bot}$ the transverse gauge field. 
The light-front expression of $g_T$ makes the physical picture clear: 
it explicitly shows the contributions associated with the quark mass, 
quark transverse momentum and quark-gluon coupling operators. 

Now we calculate $g_1$ and $g_T$ in single quark states. Here the 
calculation is performed in the light-front time-ordering perturbation
theory, in which the single quark state with fixed helicity can 
be expressed as
\begin{eqnarray}
	|k^+,k_\bot,\lambda \rangle &=& {\cal N}\Bigg\{ b^\dagger_\lambda
		(k)|0\rangle + \sum_{\lambda_1\lambda_2} \int {dk_1^+d^2
		k_{\bot 1}\over 2(2\pi)^3} {dk_2^+d^2k_{\bot 2}\over 2
		(2\pi)^3k_2^+} 2(2\pi)^3 \delta^3(k-k_1-k_2) \nonumber \\
	&&~~~~~~~~~~~~~~~~~~~~~~~~~~~~~~~\times \Phi^\lambda
		_{\lambda_1\lambda_2}(x,\kappa_\bot)b^\dagger_{\lambda_1}
		(k_1) a^\dagger_{\lambda_2} (k_2) | 0 \rangle + 
		\cdots\Bigg\}, \label{dsqs}
\end{eqnarray}
where ${\cal N}$ is the normalization constant determined by
\begin{equation}
	\langle {k'}^+,k'_\bot,\lambda' |k^+,k_\bot,\lambda \rangle
	= 2(2\pi)^3 k^+ \delta_{\lambda,\lambda'}\delta(k^+-{k'}^+)
	\delta^2(k_\bot-k'_\bot), 
\end{equation}
$b_\lambda^\dagger (k)$ and $a_\lambda^\dagger (k)$ the creation 
operators of quarks and gluons on the light-front which are
defined by \cite{Zhang93}:
\begin{eqnarray}
	\psi_+(x) &=& \sum_\lambda \chi_\lambda \int {dk^+d^2k_\bot
		\over 2(2\pi)^3}\Big(b_\lambda(k)e^{-ikx} + 
		d_{-\lambda}^\dagger(k)e^{ikx} \Big) ,  \\
	A_{a\bot}^i(x) &=& \sum_\lambda \int {dk^+d^2k_\bot\over 
		2(2\pi)^3k^+}\Big(\varepsilon^i_a(\lambda) 
		a_\lambda(k)e^{-ikx} + h.c \Big)
\end{eqnarray}
with
\begin{eqnarray} 
	\Big\{b_\lambda (k), b_{\lambda'}^\dagger(k') \Big\} &=& 
		\Big\{d_\lambda(k), d_{\lambda'}^\dagger{k'} \Big\} 
		= 2(2\pi)^3 \delta (k^+-{k'}^+) \delta^2(k_\bot - k_\bot'), \\
	\Big[a_\lambda(k) , a_{\lambda'}^\dagger (k') \Big] &=& 
		2(2\pi)^3 k^+\delta (k^+-{k'}^+) \delta^2(k_\bot - k_\bot'), 
\end{eqnarray}
and $\chi_\lambda$ is the eigenstate of $\sigma_z$ in the two-component
spinor of $\psi_+$ by the use of the following light-front $\gamma$ 
matrix representation \cite{Zhang95},
\begin{equation}
	\gamma^0 = \left[\begin{array}{cc} 0 & - i \\ i & 0 \end{array}
		\right] ~~, ~~ 
	\gamma^3=  \left[\begin{array}{cc} 0 & i \\ i & 0 \end{array}
		\right] ~~, ~~
 	\gamma^i =  \left[\begin{array}{cc} -i\tilde{\sigma}^i & 0 \\ 
		0 & i\tilde{\sigma}^i \end{array} \right] 
\end{equation}
with $\tilde{\sigma}^1 =\sigma^2, \tilde{\sigma}^2=-\sigma^1$) and 
$\varepsilon^i_a(\lambda)$ the polarization vector of transverse 
gauge field. The amplitude $\Phi^\lambda_{\lambda_1\lambda_2}
(x,\kappa_\bot)$ in eq.(\ref{dsqs}) is 
\begin{eqnarray}
	\Phi^\lambda_{\lambda_1\lambda_2}(x,\kappa_\bot) &=& - g
		T^a {x(1-x)\over \kappa_\bot^2 + m_q^2(1-x)^2}
		\chi^\dagger_{\lambda_1} \Bigg\{2{\kappa_\bot^i \over
		1-x} \nonumber \\
	&& ~~~~~~~~~~~~~ + {1\over x}(\tilde{\sigma_\bot}\cdot \kappa_\bot)
		\tilde{\sigma}^i -im_q\tilde{\sigma}^i{1-x\over x}\Bigg\}
		\chi_\lambda \varepsilon_a^{i*}(\lambda_2) . \label{ap}
\end{eqnarray}
Note that the $m_q$ dependence in the above wave function has arisen 
from the helicity flip part of light-front QCD Hamiltonian. This is
an essential term in the determination of transverse polarization
dynamics. The transverse polarized quark target in the $x$-direction 
can then be expressed by 
\begin{equation}  \label{thb}
	| k^+, k_\bot, S^1 \rangle = {1\over \sqrt{2}}\Big(| k^+, 
		k_\bot, \uparrow \rangle \pm | k^+, k_\bot, \downarrow 
		\rangle \Big)
\end{equation}
with $S^1=\pm m^R_q$, and $m^R_q$ is the renormalized quark mass.

Without the QCD correction (i.e., for the free quark state), it 
is easy to show that  
\begin{equation}
	g_T(x) =g_T^m(x) ={e_q^2\over 2} {m_q \over S^1} \delta(1-x)
		= {e_q^2\over 2} \delta(1-x), 
	~~~ g_T^{k_\bot}(x)=0=g_T^g(x).
\end{equation}
Here $m_q/S^1 = 1$ since the renormalized mass is the same as the 
bare mass at the tree level of QCD. We see that only the quark mass 
term contributes to $g_T$ in eq.(\ref{gtlf}). The quark transverse
momentum term alone does not contribute to $g_T$ since it cannot
cause helicity flip in the free theory. This result indicates that 
physically the dominant contributions to $g_T$ is not controlled 
by the twist classification. A direct calculation of $g_1(x)$ from 
eq.(\ref{g1lf}) in the free quark helicity state will immediately 
lead to the well-known solution:
\begin{equation}
	g_1(x) =  {e_q^2\over 2} \delta(1-x) .
\end{equation}
Thus, for the free quark, we have 
\begin{equation}  \label{fqg2}
	g_2(x)=g_T(x) - g_1(x) = 0 .
\end{equation} 

It is obvious that for free theory, the Burkhardt-Cottingham
(BC) sum rule is trivially obeyed. But as we can see (as it has also 
been previously noticed) the Wandzura-Wilczek relation,
\begin{equation}
	g_2(x) = - g_1(x) + \int_x^1 dy {g_1(y)\over y},
\end{equation}
is not satisfied in free theory. Jaffe and Ji argued that the 
WW relation is violated in free theory because the quark mass 
contribution in $g_2(x)$ which is thought to be a twist-3 
contribution \cite{Jaffe91} cancels the $g_2^{ww}$ contribution 
(a twist-2 contribution). This has become a common understanding
why WW relation is not satisfied in free theory. However, this may 
not be a proper explanation because if one ignored the quark mass 
contribution (naively started with a massless quark theory), then 
there would be no transverse polarization. In other words, one 
cannot predefine $g_T(x)$ in a massless quark theory. Thus the 
investigation of $g_2$ in terms of a massless quark theory is 
totally ambiguous in this way. On the other hand, if one started 
with a massive quark theory, the resulting $g_T$ would be independent of 
quark mass (the mass dependence does not occur for a quark target). 
One can then take the chiral limit ($m_q =0$) at the end of 
calculation to obtain the unambiguous $g_T$ in the massless 
theory which is the same as that in the massive theory. Thus, 
there is no so-called cancellation between the twist-two 
contribution and the quark mass contribution in the determination 
of $g_2$ in the free theory. Eq.(\ref{fqg2}) just indicates that 
there does not exist the WW relation, at least in free theory. 
Such a picture is not  manifest in OPE approach.

Next, we consider the QCD corrections up to order $\alpha_s$, where
the quark-gluon interaction is explicitly included.  We find that all 
the three terms in eq.(\ref{gtlf}) have nonzero contribution 
to $g_T$, 
\begin{eqnarray}
	g_T^m(x,Q^2) &=&{e^2_q\over 2} {m_q\over S^1}
 		\Bigg\{\delta(1-x) + {\alpha_s \over 2\pi} 
		C_f \ln{Q^2\over \mu^2} \Bigg[{2 \over 
		1-x} \nonumber \\
	&& ~~~~~~~~~~~~~~  -~\delta(1-x) \int_0^1 dx'
		{1+x'^2\over 1-x'} \Bigg]\Bigg\}, \label{ems}\\
	g_T^{k_\bot}(x,Q^2) &=&-{e^2_q\over 2} {m_q\over S^1} 
		{\alpha_s \over 2\pi} C_f \ln{Q^2\over \mu^2}(1-x) , 
		\label{gtk} \\
	g_T^g(x,Q^2) &=& {e^2_q\over 2} {m_q\over S^1}{\alpha_s 
		\over 2\pi} C_f \ln{Q^2\over \mu^2}{\delta(1-x) \over 2}, 
		\label{gtg}
\end{eqnarray}
where we have set a hadronic scale such that $|k_\bot|^2 >> \mu^2 >> 
(m_q)^2 $ and $\mu$ is the hadronic factorization scale for separating 
the ``hard" and ``soft" dynamics of QCD. As a matter of fact, the 
above result represents
purely the pQCD dynamics. The integral term in eq.(\ref{ems}) comes 
from the dressed quark wavefunction normalization constant ${\cal N}$ 
in eq.(\ref{dsqs}) (corresponds to the virtual contribution in the 
standard Feynman diagrammatic approach). It shows that up to the 
order $\alpha_s$, the matrix elements from $O_{k_\bot}$ (quark 
transverse momentum effect) and $O_g$ (quark-gluon interaction
effect) in eq.(\ref{gtlf}) are also proportional to quark mass. 
In other words, the transverse quark momentum and quark-gluon 
coupling contributions to $g_T(x,Q^2)$ arise from quark mass effect. 
Explicitly, these contributions arise from the interference of the 
$m_q$ term with the non-$m_q$ dependent terms in the wave function
of eq.(\ref{ap}) through the quark transverse momentum operator
and the quark-gluon coupling operator in the $g_T$ expression.
This result is not surprising since, as we have pointed out, the 
pure transverse polarized structure function measures the dynamical 
effect of chiral symmetry breaking \cite{Zhang96}. Physically 
only these interferences related to quark mass can result in
the helicity flip (i.e., chiral symmetry breaking) in pQCD so 
that they can contribute to $g_T(x,Q^2)$. From this result, we 
may see that only the operators themselves or their twist structures 
may not give us useful information about their importance in 
the determination of structure functions.

Combining the about results together, we obtain 
\begin{equation}
	g_T(x,Q^2) = {e^2_q \over 2} {m_q \over S^1} \Bigg\{
		\delta(1-x) + {\alpha_s \over 2\pi} C_f \ln{Q^2\over 
		\mu^2} \Bigg[{1+2x-x^2 \over (1-x)_+} + 2 \delta(1-x) 
		\Bigg] \Bigg\} . \label{gt1}
\end{equation}
Note that in the above solution, $m_q$ is the bare quark mass,
while the dressed quark polarization $S^1= m^R_q$, and
up to order $\alpha_s$, 
\begin{equation}
	m_q^R = m_q \Bigg( 1 + {3\over 4\pi} \alpha_s C_f \ln
		{Q^2\over \mu^2} \Bigg) .
\end{equation}
We must emphasize that on the light-front there are two mass
scales in the QCD Hamiltonian, one is proportional to $m_q^2$
which does not violate chiral symmetry, and the other is 
proportional to $m_q$ which we discuss here and is associated
with explicit chiral symmetry breaking in QCD \cite{Wilson94}.
 An important
feature of light-front QCD is that the above two mass scales
are renormalized in different ways even in the perturbative 
region.  The renormalization of $m_q^2$ in pQCD is different
from the above result, the details of which can be found in
 our previous work \cite{Zhang93}. 
With this consideration, we have
\begin{equation}
	g_T(x,Q^2) = {e^2_q\over 2}  \Bigg\{\delta(1-x) + 
		{\alpha_s \over 2\pi} C_f \ln{Q^2\over \mu^2} 
		\Bigg[{1+2x-x^2 \over (1-x)_+} + {1\over 2} 
		\delta(1-x) \Bigg] \Bigg\} . \label{gts}
\end{equation}
The final result is independent quark mass but we have to emphasize
again that one must start with massive quark theory. Otherwise, 
there is no definition for $g_T$ at the beginning. 
 
By a similar calculation for $g_1$,  we have
\begin{equation}  \label{g1s}
	g_1(x,Q^2) = {e^2_q \over 2} \Bigg\{\delta(1-x) + {\alpha_s 
		\over 2\pi} C_f \ln{Q^2\over \mu^2} \Bigg[{1+x^2 
		\over (1-x)_+} + {3\over 2}\delta(1-x) \Bigg]\Bigg\} ,
\end{equation}
which is independent of quark mass in any step of calculation. Thus, 
up to of the oder $\alpha_s$, we find $g_2$ for a quark target 
\begin{equation}
	g_2(x,Q^2) = {e_q^2\over 2}{\alpha_s \over 2\pi} C_f 
		\ln{Q^2\over \mu^2} \Big[2x - \delta(1-x) \Big].
\end{equation}
It is easy to check that the above result of $g_2(x,Q^2)$ obeys 
the BC sum rule,
\begin{equation}
	\int_0^1 dx g_2(x,Q^2) = 0,
\end{equation}
as is expected. However, it also shows that the WW relation is 
violated. It is not possible to let $g_2$ satisfy the WW relation 
by ignoring quark mass effect and/or quark-gluon interaction
contributions in pQCD. In the physical space, one cannot separate 
quark mass contribution and quark-gluon coupling contribution
and ignore the quark mass effect in addressing the transverse
polarization processes. 

\vspace{0.3in}

\noindent{\bf 3. Comparison with the OPE formulation}

As is well-known, the WW relation was originally derived from OPE 
method by ignoring the twist-3 contribution to $g_2(x,Q^2)$. The 
OPE method has been thought as a fundamental approach in the 
investigation of DIS structure functions. Using the OPE, one can 
show that the moments of the polarized structure functions $g_1$ and 
$g_2$ can be expressed by 
\begin{eqnarray}
	&& \int_0^1 dx x^{n-1} g_1(x,Q^2) = {1\over 2} a_n , ~~~
		n = 1, 3, 5, ..., \label{opeg1} \\
	&& \int_0^1 dx x^{n-1} g_2(x,Q^2) = -{1\over 2} {n-1\over n}
		(a_n - d_n) , ~~~n = 3, 5, ...,  \label{opeg2}
\end{eqnarray}
where $a_n$ and $d_n$ are determined by the twist-2 and twist-3
hadronic matrix elements. Note that OPE is incomplete in describing
$g_2$ since it cannot define the first moment of $g_2$. In fact,
the OPE only defines the odd moments of the polarized structure 
functions.

From eqs.(\ref{opeg1}-\ref{opeg2}),
\begin{equation}
	\int_0^1 dx x^{n-1} \Big\{g_1(x,Q^2) + g_2(x,Q^2) \Big\} 
		= {1\over 2n} a_n  +{1\over 2} {n-1\over n} d_n ,
		 ~~~n = 3, 5, ...,  \label{opegt}
\end{equation}  
If one {\it assumed} that eq.(\ref{opegt}) is valid for all integers $n$,
then using the convolution theorem for Mellin transformation,
one could find
\begin{equation}
	g_2(x,Q^2) = - g_1(x,Q^2) + \int_x^1 {dy \over y} g_1(y,Q^2)
		+ \bar{g}_2(x,Q^2)	, \label{opef}
\end{equation}
where $\bar{g}_2(x,Q^2)$ is the so-called twist-3 contribution
that is related to $d_n$.
If one further {\it assumed} that the twist-3 contribution $\bar{g}_2$
can be ignored, then eq.(\ref{opef}) would be reduced to $g_2^{WW}(x,Q^2)$
that obeys the famous WW relation. Obviously, to obtain the WW relation,
one has to make some unjustified assumptions.

Now we shall examine whether one can assume the validity of eq.(\ref{opegt}) 
for all integers $n$ and how big the twist-3 contribution can be in pQCD.

In our approach, instead of the moments, we directly calculate the structure
functions $g_1$ and $g_2$. Then using the identity,
\begin{equation}
	\int_0^1 dx x^{n-1} {1\over (1-x)_+} = 
		- \sum_{j=1}^{n-1} {1\over j} ,
\end{equation}
we calculate from eqs.(\ref{g1s}-\ref{gts}) all 
the moments for $g_1(x,Q^2)$ and $g_T(x,Q^2)$ up to $\alpha_s$
in pQCD. The result is 
\begin{eqnarray}
	&& \int_0^1 dx x^{n-1} g_1(x,Q^2)  
		= {e_q^2\over 2} \Bigg\{ 1 + {\alpha_s \over 2\pi}
		\ln{Q^2\over \mu^2}\Bigg[-{1\over 2} +{1\over n(n+1)}
		- 2\sum_{j=2}^n {1\over j} \Bigg] \Bigg\} ,  \label{lfg1} \\
	&& \int_0^1 dx x^{n-1} g_T(x,Q^2)  
		= {e_q^2\over 2} \Bigg\{ 1 + {\alpha_s \over 2\pi}
		\ln{Q^2\over \mu^2}\Bigg[-{3\over 2} +{1\over n}
                               +{1 \over n+1}
		- 2\sum_{j=2}^n {1\over j} \Bigg] \Bigg\} .  \label{lfgt}
\end{eqnarray}
Since the above result is directly obtained from pQCD without using
the OPE, it is valid for all the integers $n$. It is easy to see
that the first moment of $g_1$ is the same as that of $g_T$ for
quark target,
\begin{equation}
	\int_0^1 dx g_1(x,Q^2) = \int_0^1 dx g_T(x,Q^2) = {e_q^2\over 2},
\end{equation}
and is independent on $Q^2$, as is expected in the leading order pQCD.
Again from the above result, we see that the BC sum rule is satisfied 
in pQCD.

Comparing eqs.(\ref{lfg1}-\ref{lfgt}) with eqs.(\ref{opeg1}-\ref{opegt}), 
we obtain 
\begin{eqnarray}
	&& a_n = e_q^2 \Bigg\{ 1 + {\alpha_s \over 2\pi}
		\ln{Q^2\over \mu^2}\Bigg[-{1\over 2} +{1\over n(n+1)}
		- 2\sum_{j=2}^n {1\over j} \Bigg] \Bigg\} , \\
	&& d_n = e_q^2 \Bigg\{ 1 + {\alpha_s \over 2\pi}
		\ln{Q^2\over \mu^2}\Bigg[-{3\over 2} +{1\over n}
		- 2\sum_{j=2}^n {1\over j} \Bigg] \Bigg\}.
\end{eqnarray}
The above result is valid for all the integers $n$. In other words,
when $g_2(x,Q^2)$ is written by eq.(\ref{opef}), $a_n$ and $d_n$ 
are the same order for all the $n$, \begin{equation}
	a_n - d_n =  e_q^2 {\alpha_s \over 2\pi} \ln{Q^2\over \mu^2}
		\Bigg[ 1 - {1\over n+1} \Big ]
\end{equation}
This shows that $\bar{g}_2(x,Q^2)$ is comparable with $g^{WW}_2(x,Q^2)$
so that the WW relation is badly violated not only in free theory but 
also in pQCD by the quark mass effects. It is easy to check that in
terms of moments, the result presented here is consistent with the 
result by Altarelli et al. \cite{Alta94} except for a finite part which 
we did not include here. This finite contribution comes from the 
region $ 0 \leq |k_\bot|^2 \leq \mu^2$ [see the discussion after 
eq.(\ref{gtg})] which is the soft parton domain, and is 
not part of pQCD contribution in the standard physical picture. But it is
straightforward to include this contribution and the result does
not change the conclusion in this paper.

It is worth noticing that here $d_n$ only represents a part of twist-3 
contribution in the real hadronic process that is associated with 
quark mass effect (including a part of contributions from quark-gluon
coupling and quark transverse momentum operators).  The possible  
contributions purely related to the quark-gluon coupling and intrinsic 
transverse quark momentum dynamics do not occur in the case of using 
quark target. Is there any possibility that these possible twist-3 
contributions in nucleon targets may cancel the quark mass effect so 
$a_n$ can become dominant in hadronic processes? None can provide a 
definite answer for this possibility at the current status of QCD. 
However, as we have pointed out recently\cite{Zhang96}, based
on the hadronic bound state structure, the pure twist-3 contribution 
related to the direct quark-gluon coupling dynamics and the quark 
transverse momentum effect is rather small since it is proportional
to the off-diagonal hadronic matrix elements in different Fock states.
Meanwhile, the pure mass effect in pQCD may also be suppressed by the
factor ${m_q \over S_\bot} \rightarrow {m_q \over M}$ in the real
hadronic process, where $M$ is the hadronic mass. However, we find
that the physical picture of the transverse polarized structure 
function is that it essentially describes the chiral symmetry 
breaking dynamics in DIS. Accompanying with the above mass effect, 
there is a nonperturbative QCD effect at the factorization scale 
that is generated by the dynamical chiral symmetry breaking. 
This effect cannot be suppressed by the factor ${1\over M}$. The 
physics of the structure function $g_2$ is not described by the WW 
relation, nor it measures the higher-twist dynamics. It has a simple 
parton picture and is characterized by the chiral symmetry breaking 
dynamics that is not described by the WW relation \cite{Zhang96}.

In conclusion, since it was proposed twenty years ago, the WW 
relation has not been seriously examined in pQCD. In this paper 
we explicitly show how the WW relation is badly violated even 
in pQCD to the order $\alpha_s$.  It is hard to think how it 
can be a good representation of $g_2$ if it cannot be satisfied 
in free theory and in pQCD.

A.H would like to thank Prakash Mathews and K. Sridhar, and W.M.Z
would like to thank H. Y. Cheng, C. Y. Cheung and H. L. Yu for 
useful discussions. This work is partially supported by 
NSC86-2112-M-001-020 and NSC86-2816-M001-009R-L(WMZ).

\end{document}